\documentclass[twocolumn,prb,aps,superscriptaddress,showpacs,amsmath,amssymb]{revtex4-1}
\usepackage{amsfonts}

\usepackage{graphicx}
\usepackage{dcolumn}
\usepackage{bm}
\usepackage{amsmath}
\usepackage{amssymb}
\usepackage{subfigure}
\usepackage{xcolor}
\usepackage[colorlinks]{hyperref}

\renewcommand\P{P_{\rm S}}
\newcommand\up{\uparrow}
\newcommand\down{\downarrow}
\newcommand\Right{{\rm R}}
\newcommand\Left{{\rm L}}

\begin{document}

\title{Contact effects in spin transport along double-helical molecules}
\author{Ai-Min Guo}
\thanks{These authors contributed equally to this work.}
\affiliation{Institute of Physics, Chinese Academy of Sciences, Beijing 100190, China}
\author{E. D\'{i}az}
\thanks{These authors contributed equally to this work.}
\affiliation{GISC, Departamento de F\'isica de Materiales, Universidad Complutense, E-28040 Madrid, Spain}

\author{C. Gaul}
\affiliation{Max Planck Institute for the Physics of Complex Systems, 01187 Dresden, Germany}

\author{R. Gutierrez}
\affiliation{Institute for Materials Science, Dresden University of Technology, 01062 Dresden, Germany}

\author{F. Dom\'{i}nguez-Adame}
\affiliation{GISC, Departamento de F\'isica de Materiales, Universidad Complutense, E-28040 Madrid, Spain}

\author{G. Cuniberti}
\affiliation{Institute for Materials Science, Dresden University of Technology, 01062 Dresden, Germany}
\affiliation{Center for Advancing Electronics Dresden, TU Dresden, 01062 Dresden, Germany}
\affiliation{Dresden Center for Computational Materials Science, TU Dresden, 01062 Dresden, Germany}

\author{Qing-feng Sun}
\affiliation{International Center for Quantum Materials, School of Physics, Peking University, Beijing 100871, China}
\affiliation{Collaborative Innovation Center of Quantum Matter, Beijing 100871, China}

\begin{abstract}

We report on spin transport along double-helical molecular systems by considering various contact configurations and asymmetries between the two helical strands in the regime of completely coherent charge transport. Our results reveal that no spin polarization appears in two-terminal molecular devices when coupled to one-dimensional electrodes. The same holds in the case of finite-width electrodes if there is a \emph{bottleneck\/} of one single site in the system electrode--molecule--electrode. Then, additional dephasing is necessary to induce spin-filtering effects. In contrast, nonzero spin polarization is found in molecular devices with multiple terminals or with two finite-width electrodes, each of them connected to more than one site of the molecule. Then, the magnitude of spin polarization can be enhanced by increasing the asymmetry between the two strands. We point out that the spin-filtering effects could emerge in double-helical molecular devices at low temperature without dephasing by a proper choice of the electrode number and the connection between the molecule and the electrodes.

\end{abstract}


\maketitle

\section{Introduction}

Recent experiments on spin transport along DNA molecules have made an important breakthrough in molecular spintronics.\cite {goehler11,xie,kumar} G\"{o}hler \textit{et al.\/} reported that the electrons transmitted through self-assembled monolayers of double-stranded DNA (dsDNA) at room temperature were highly polarized and that the spin polarization increased with the DNA length, while spin-dependent effects were not observed in single-stranded DNA (ssDNA) monolayers.\cite{goehler11} Later on, by measuring charge transport in a two-terminal setup, Xie \textit{et al.}\ further substantiated that single dsDNA molecules can act as very efficient spin filters.\cite{xie} Very recently, Mishra \textit{et al.\/} also found spin-dependent effects in the system of single-helical bacteriorhodopsin on gold and aluminum substrates.\cite{zach2013} These experimental works have been generating intense interest in the organic spintronics based on helical molecules.

From the theoretical side, Guo and Sun\cite{Guo1,Guo2,Guo3} and Gutierrez \textit{et al.\/}\cite{Gutierrez12,Gutierrez13} proposed minimal model Hamiltonian approaches to rationalize spin selectivity. Both approaches relied on the basic assumption that a helical electric field, mirroring the helical symmetry of the molecular systems, could be the source of a spin-orbit coupling (SOC) contribution, thus relating spin propagation and helical symmetry.

While the above theoretical approaches were more appropriate for describing a charge or spin \textit{transport} set-up, earlier
investigations addressed the problem from the point of view of scattering theory, which is more suitable for the photoemission
experiments,\cite{goehler11} where the moving charges are not probing the molecular orbital structure of the system due to their
higher energies (above the vacuum level). Thus, Yeganeh \textit{et al.} studied the  transmission of spin-polarized electrons through helically shaped potentials,\cite{yeganeh} but found a very weak effect for realistic model parameters. These studies were extended in Ref.~\onlinecite{medina} to include decoherence effects within the framework of scattering theory.

More recently, Gersten \textit{et al.} suggested that the role of SOC in the metallic substrate should also be taken into
account.\cite {gersten} They showed that an interplay of strong substrate SOC and molecular helical symmetry could also induce
spin selectivity. Eremko and Loktev considered an incoming electron in a pure spin state and its reflection and transmission across a helical potential; they analytically provided the spin polarization of the transmitted electron.\cite{Eremko2013} We also mention studies by Vager and Vager relating spin-dependent effects to the presence of bound states in charge motion along a curved path.\cite{vager} Independently of the specific details of the proposed models and of the experimental setups, all studies listed above clearly agree in suggesting that the helical symmetry of the probed molecules is a key ingredient in inducing spin-selective electron transmission.\cite{naaman12}

Turning again to the electrical transport experiments,\cite{xie} it is well-known from the field of molecular electronics that the molecule-electrode interface plays a critical role in determining the electrical response of the system (see, e.g. Ref.~\onlinecite{tao1}). Generally speaking, the contacts can be classified into two categories. Firstly, a contact can be realized through physical adsorption of the molecule to the electrodes (physisorption). For instance, the two terminals of the molecule can be directly deposited on the electrodes,\cite{fink,pablo} or they can be attracted to the nanoparticles of the metal electrodes by electrostatic trapping.\cite{bezryadin,porath} In this case, several sites of the molecule may be attached to the electrodes with finite cross-section. Secondly, the contact can be also achieved through chemisorption between the molecule and the electrodes mediated e.g., by thiol groups.\cite{cui,xu,cohen,guo} Such chemical contacts favor reproducible results and may not mask the intrinsic switching properties of the molecule.\cite{tao1} In this second category, there is usually one site of the molecule connected to the electrodes.

Motivated by these issues, we explore in this paper the spin transport properties of \textit{double-helical}~(DH) molecular systems by considering various contact configurations in the fully coherent charge transport regime. Our results show a fundamental distinction for different contacts. The spin polarization turns out to be exactly zero in two-terminal molecular devices when they are coupled to one-dimensional (1D) electrodes or if only a single site at each edge of the molecule is attached to an electrode with finite cross-section. In this case, dephasing is a necessary ingredient to induce the spin-filtering effects. Contrarily, nonzero spin polarization appears in a multi-terminal set-up, or a two-terminal one with more than one site of the molecule connected to each finite-width electrode. Additionally, we also investigate the influence of the asymmetry between the two helical chains of the molecule and find that the magnitude of spin polarization could be enhanced by increasing this asymmetry. We remark that these results are consistent with the main conclusions in Refs.~\onlinecite{Guo1} and \onlinecite{Gutierrez13}. We finally point out that the spin-filtering effects could emerge in the DH molecular devices, e.g., dsDNA devices, at low temperature without dephasing by properly tuning the electrode number and the connection between the molecule and the electrodes.

The rest of the paper is organized as follows. In Sec. II, the theoretical model and the parameters are presented.
In Secs. III and IV, we investigate the conductance and the spin polarization for the two- and the multi-terminal set-up with
1D electrodes. The study of the two-terminal set-up with finite-width electrodes is found in Sec. V and its particular
situation with bottleneck in Sec. VI. Finally, a brief summary is presented in Sec. VII.

\section{Model and parameters}

Spin transport along the DH molecule can be described by the Hamiltonian ${\cal H}={\cal H}_{\rm m}+{\cal H}_{\rm ec}$, where ${\cal H}_{\rm m}$ and ${\cal H}_{\rm ec}$ describe the DH molecule and the electrodes including the molecule-electrode coupling, respectively. The DH molecule is represented by a two-leg ladder model \cite{grozema,wang} including the SOC term and can be expressed as\cite{Guo1}
\begin{eqnarray}
{\cal H}_{\rm m}&=&\sum_{j=1}^{2} \Big\{\sum_{n=1}^{N} \varepsilon_{j} c_{jn}^{\dagger}c_{jn}^{} \!+\! \sum_{n=1}^{N-1} \!\Big[ i \gamma_{j} c_{jn}^{\dagger}(\sigma_n^{(j)} + \sigma_ {n+1}^{(j)})c_{jn+1}^{} \nonumber \\ &+& t_{j}c_{jn}^{\dagger}c_{jn+1}^{} + {\rm H.c.} \Big]\Big\} + \sum_{n=1}^{N} \Big[\lambda c_{1n}^{\dagger}c_{2n}^{}+ {\rm H.c.} \Big]\ , \label{eq1}
\end{eqnarray}
where $c_{jn}^{\dagger}= (c_{jn\uparrow}^\dagger, c_{jn\downarrow} ^\dagger)$ and  $c_{jn}= (c_{jn\uparrow}, c_{jn\downarrow})^ \intercal$ are, respectively, the creation and annihilation operators at site $\{j,n\}$ of the DH molecule whose length is $N$. Here, $j$ labels the helical chain, $n$ is the site index within a single chain, and $^\intercal$ means the transpose. $\varepsilon_j$ is the on-site energy, $\gamma_j$ is the SOC strength, and $t_j$ ($\lambda$) is the intrachain (interchain) hopping integral. Finally, the term $\sigma_{n+1}^{(j)}=\sigma_z \cos\theta - (-1)^j \sin\theta [\sigma_x \sin (n\Delta\varphi) - \sigma_y \cos (n\Delta\varphi)]$ with $\sigma_{x,y,z}$ the Pauli matrices, $\theta$ the helix angle, and $\Delta\varphi$ the twist angle between successive base pairs.\cite{Guo1} Note that if the factor $(-1)^j$ is dropped in the expression of $\sigma_{n}^{(j)}$, the electronic states along the two helical strands $j=1,2$ of the Hamiltonian \eqref{eq1} can alternatively be interpreted as being two different electronic states along a single-stranded molecule. In this case, this model would be similar to that formulated in Ref.~\onlinecite{Gutierrez13}. We remark however, that the factor $(-1)^j$ changes the symmetry of the system and thus the behavior of the spin polarization is not exactly the same.

For the sake of clarity, the electronic parameters are regarded as uniform along each helical chain of the DH molecule. Here, let us focus on the DNA molecule as an example. It is well established that the double-helical structure of the DNA molecule is constructed by four nucleobases---guanine (G), adenine (A), cytosine (C), and thymine (T)---based on the complementary base-pairing rule. Since both the structures and the number of atoms of these nucleobases are different from each other, there may exist asymmetries in the electronic parameters between the two DNA strands, as demonstrated from first-principles calculations.\cite{voityuk,senthilkumar,hawke}  Therefore, we employ an additional parameter $x$ to describe this asymmetry. Some of the electronic parameters are fixed to be $\varepsilon_1=0$, $\varepsilon_2=0.3$, $t_2=0.1$, $\lambda=-0.08$, and $\gamma_1=0.01$, in units of eV, while others are taken as $t_1=-x t_2$ and $\gamma_2=x\gamma_1$ to simulate the asymmetry between the two helical chains. The remaining parameters are chosen as $N=30$, $\theta=0.66$ rad, and $\Delta\varphi=\pi/5$, resembling the B-form dsDNA molecule. Notice that the asymmetry parameter $x$ is included in an inverse way between the intrachain hopping integrals and the SOCs. This parametrization was demonstrated as the favorable situation to obtain large spin polarization.\cite{Gutierrez13}

The remaining Hamiltonian ${\cal H}_{\rm ec}={\cal H}_{\rm e}^{(D,J)}+ {\cal H}_{\rm c}$ splits into the Hamiltonian of the left and right nonmagnetic electrodes ${\cal H}_{\rm e}^{(D,J)}$ and the Hamiltonian ${\cal H}_{\rm c}$ for the coupling between the molecule and the electrodes. Here, $D$ denotes the number of legs of the electrodes, while $J$ refers to the number of electrodes connected to each end of the molecule. The Hamiltonians ${\cal H}_{\rm e}^{(D,J)}$ and ${\cal H}_{\rm c}$ depend on the particular realization of the electrodes and the connection between the molecule and the electrodes, which has an enormous impact on the spin transport along the DH molecule as discussed below.

Spin transport will be studied by considering the Landauer-B\"{u}ttiker formalism. Thus, the conductance with spin $s=\uparrow,\downarrow$ in the right electrode is\cite{Ryndyk2009}
\begin{eqnarray}
G_s= \frac{e^2}{h}\, {\rm Tr} \big[{\bm \Gamma}_{\Right s}{\bm G}^{\rm r} {\bm \Gamma}_\Left {\bm G}^{\rm a}\big ]\ . \label{eq2}
\end{eqnarray}
Here, ${\bm \Gamma}_{\ell} =i({\bm \Sigma}_{\ell}^{\rm r} - {\bm \Sigma}_{\ell}^{\rm a})$, with $\ell=\Left,\Right s$, is the linewidth function, ${\bm G}^{\rm r}=({\bm G}^{\rm a})^\dagger = (E {\bm I} - {\mathcal H}_{\rm m} -{\bm \Sigma}_\Left^{\rm r}-  {\bm \Sigma}_{\Right\uparrow}^{\rm r}-{\bm \Sigma}_{\Right\downarrow}^ {\rm r})^{-1}$ is the Green's function, ${\bm \Sigma}_{\ell}^{\rm r}$ is the retarded self-energy due to the coupling to the left or right electrode. Then, the spin polarization is defined as
\begin{equation}
\P=(G_{\uparrow} - G_{\downarrow})/(G_{\uparrow}+G_{\downarrow})\ ,
\label{eqPolarization}
\end{equation}
which is the physical quantity of interest hereafter.

\begin{figure}[ht]
\includegraphics[width=0.4\textwidth]{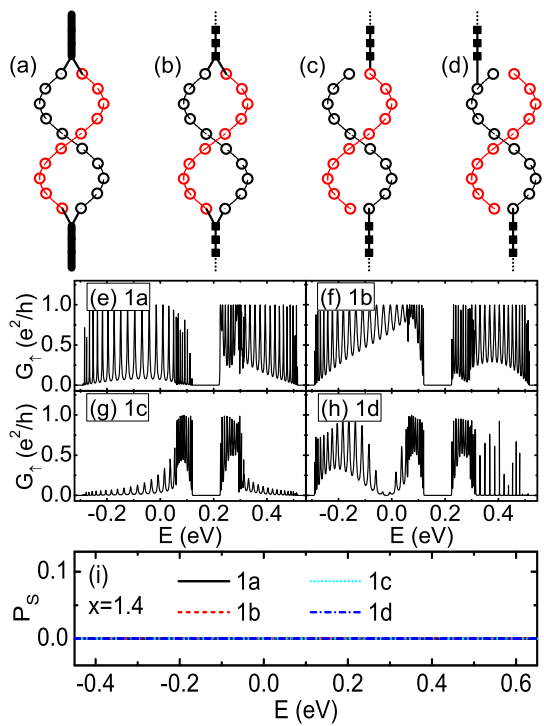}
\caption{\label{fig:one}(Color online) (a)--(d) Schematic views of various two-terminal DH molecular devices coupled to 1D electrodes. Each circle denotes an atom or an atomic cluster, with the black (red) ones assembling the first (second) helical chain. In these figures, although the electrodes locate on the bottom and top sides of the DH molecule, we still call them the left and right electrodes, respectively, in the text. (e)--(h) Energy-dependent spin-up conductance $G_\uparrow$ and (i) spin polarization $\P$ with $x=1.4$.}
\end{figure}

\section{Two-terminal set-up with 1D electrodes}

Figures~\ref{fig:one}(a) and \ref{fig:one}(b)--\ref{fig:one}(d) show the two-terminal DH molecular devices coupled to the 1D normal metal electrodes in continuous and discrete forms, respectively. In the following, each molecular device will be named by the number and label of the figure where it is represented, e.g., the molecular device in Fig.~\ref{fig:one}(a) is denoted as model 1a. The Hamiltonian for the discrete electrodes is written in real space as
\begin{eqnarray}
{\cal H}_{\rm e}^{(1,1)} \!=\! \! \sum_{n=-\infty}^{-1} t_0 a_{n}^\dagger a_{n+1}^{} +\! \! \sum_{n=N+1}^{+\infty} t_0 a_{n}^\dagger a_{n+1}^{} + {\rm H.c.}\ , \label{eqHamiltonianDiscrete}
\end{eqnarray}
which represents the 1D semi-infinite electrodes with the on-site energy being zero and the hopping integral $t_0$. In continuous case 1a, we resort to the momentum space representation\cite{Guo1}
\begin{equation}
{\cal H}_{\rm ec}^{\rm (1a)} \! =\!\!\!\! \sum_{k,\beta=\Left, \Right} \left[ \varepsilon_{\beta k}a_{\beta k}^\dagger a_{\beta k}^{} + t_\beta a_{\beta k}^\dagger (c_{1n_\beta}^{}+ c_{2n_\beta} ^{}) + {\rm H.c.} \right]\ , \label{eq3}
\end{equation}
where $a_{\beta k}^\dagger=(a_{\beta k\uparrow}^\dagger,a_{\beta k\downarrow}^\dagger)$ is the creation operator of mode $k$ in electrode $\beta$, $t_{\beta}$ is the coupling between the molecule and the left/right electrode, $n_\Left=1$, and $n_\Right=N$. Both helical chains are attached to the left (right) electrode at sites $\{1,1\}$ and $\{2,1\}$ ($\{1,N\}$ and $\{2,N\}$). In the numerical calculations, the linewidth functions are assumed to be energy-independent (wide-band limit) and set to ${\bf \Gamma}_{\beta} =2\pi\rho_{\beta} t^2_{\beta} =1$, where $\rho_{\beta}$ refers to the density of states of the electrodes.\cite{Guo1}

In all three discrete models 1b--1d, the Hamiltonian ${\cal H}_{\rm e} ^{(1,1)}$ is given by Eq.~\eqref{eqHamiltonianDiscrete}. However, different sites of the DH molecule are attached to the electrodes in these models: In 1b, both strands are coupled to the electrodes; in 1c every single chain is attached to one of the two electrodes (sites $\{1,1\}$ and $\{2,N\}$ are coupled to the left and right electrodes, respectively); in 1d the first helical chain is attached to both electrodes, namely at sites $\{1,1\}$ and $\{1,N-1\}$.  The coupling Hamiltonians for each model read
\begin{align}
{\cal H}_{\rm c}^{\rm (1b)}&=t_\Left a_{0}^\dagger (c_{1 1}^{}+c_{2 1}^{}) + t_\Right a_{N+1}^\dagger (c_{1 N}^{}+c_{2 N}^{}) + {\rm H.c.}\ , \nonumber\\ {\cal H}_{\rm c}^{\rm (1c)}&=t_\Left a_{0}^ \dagger c_{1 1}^{} + t_\Right a_{N+1}^\dagger c_{2 N}^{} + {\rm H.c.}\ , \nonumber\\ {\cal H}_{\rm c}^{\rm (1d)}&=t_\Left a_{0}^ \dagger c_{1 1}^{} + t_\Right a_{N+1}^\dagger c_{1 N-1}^{} + {\rm H.c.}\ . \label{eq5}
\end{align}
In this situation, the retarded self-energy can be calculated numerically \cite{Lee} and the parameters are taken as $t_0=2$ eV and $t_\Left=t_\Right=0.3$ eV hereafter.

Figures~\ref{fig:one}(e)--\ref{fig:one}(h) plot the spin-up conductance $G_\uparrow$ for models 1a--1d, respectively, with the asymmetry parameter $x=1.4$. Although all these molecular devices possess distinct contact configurations, the transmission spectra always consist of two subbands separated by a well-defined band gap.  Although we are considering single-electron physics here, we use the conventional labeling and refer to these bands as the highest occupied molecular orbital (HOMO) band and the lowest unoccupied molecular orbital (LUMO) band, respectively. In both bands, due to the SOC effects, the energy region close to the gap presents a higher density of the transmission peaks. However, the conductance is very sensitive to the contact configuration. If both helical chains are connected to the electrodes, resonant states with $G_\uparrow=1$ are found in the whole range of the bands in the energy spectrum [Figs.~\ref{fig:one}(e) and \ref{fig:one}(f)], while if only a single site of the  molecule is attached to each electrode, the resonant states are considerably suppressed [Figs.~\ref{fig:one}(g) and \ref{fig:one}(h)]. In all four cases, the conductance $G_\up$ is equal to $G_\down$. As indicated in Fig.~\ref{fig:one}(i), the corresponding spin polarization $\P$ is exactly zero for all the two-terminal molecular devices connected to the 1D electrodes, regardless of the particular way the molecule coupled to the electrodes and the asymmetry between the two helical chains. This behavior is related to the time-reversal symmetry and the phase-locking effect in two-terminal devices.\cite{Sun1,openness}

As shown in Ref.~\onlinecite{Guo1}, additional dephasing is necessary to yield non\-zero spin polarization in this situation. When an electron is transmitted through the molecular system, it may experience inelastic scattering events, which lead to the loss of phase memory, and can be simulated by attaching each site of the molecule to a B\"{u}ttiker's virtual electrode. As a result, the two-terminal devices are naturally switched into multi-terminal devices. In other words, the dephasing promotes the openness of the two-terminal devices and induces the spin-filtering effects.\cite{Guo1} Actually, the B\"{u}ttiker's virtual electrode is similar to the real one, because their Hamiltonians are identical to each other. However, with B\"{u}ttiker's probes zero current flow must be enforced through them in order to have current conservation, because the probes are not necessarily physical terminals but mathematical artefacts to induce dephasing.\cite{DAmato} Hence, one may expect that nonzero spin polarization could be observed  in the DH molecular devices when the virtual electrode is replaced by the real one, which will be investigated in the following sections.

\begin{figure}[ht]
\includegraphics[width=0.4\textwidth]{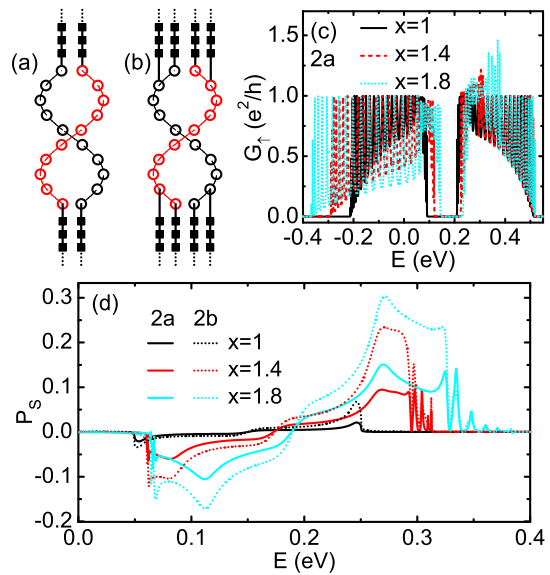}
\caption{\label{fig:two}(Color online) (a)--(b) Schematic views of the multi-terminal molecular devices. (c)~Energy-dependent $G_\uparrow$ for model 2a and (d) $\P$ for models 2a and 2b with various asymmetries.}
\end{figure}

\section{Multi-terminal set-up with 1D electrodes}

We consider now multi-terminal molecular devices by connecting the DNA molecule to several 1D semi-infinite electrodes. Figure~\ref{fig:two}(a) shows the four-terminal molecular device where each of the two helical strands is connected to a separate electrode at each end; Fig.~\ref{fig:two}(b) sketches the eight-terminal molecular device where the first and last \emph{two} sites of each strand are connected to separate electrodes. In this situation, the Hamiltonians for the electrodes in models 2a and 2b consist of $J=2,4$ copies of the electrode Hamiltonian \eqref{eqHamiltonianDiscrete}, respectively, and read
\begin{align}
{\cal H}_ {\rm e} ^{(1,J)}&= \sum_{j=1}^{J}\Big( \sum_{n=-\infty} ^{-1} t_0a_{jn}^\dagger a_{j n+1}^{}\nonumber \\ &+ \!\sum_{n=N+1 }^{+\infty}\! t_0 a_{jn}^\dagger a_{j n+1}^{}+ {\rm H.c.} \Big)\ . \label{eq6}
\end{align}
As indicated in Figs.\ \ref{fig:two}(a) and \ref{fig:two}(b), the coupling Hamiltonian $H_{\rm c}$ is given as
\begin{subequations}
\begin{eqnarray}
{\cal H}_{\rm c}^{\rm (2a)}&=&\sum_{j=1}^2 (t_\Left a_{j0}^\dagger c_{j 1}^{} + t_\Right a_{j,N+1}^\dagger c_{3-j,N}^{} + {\rm H.c.})\ , \label{eq7-1}\\ {\cal H}_{\rm c}^{\rm (2b)}&=&\sum_{j=1}^2 \Big[ t_\Right (a_{j,N+1}^\dagger c_{2, N-2+j}+a_{j+2,N+1}^\dagger c_{1,N+1-j}^{}) \nonumber \\ &+&t_\Left(a_{j0}^\dagger c_{1, 3-j}^{}+a_{2+j,0}^\dagger c_{2j})^{}+{\rm H.c.}\Big]\ . \label{eq7-2}
\end{eqnarray}
\label{eq7}
\end{subequations}

Figure~\ref{fig:two}(c) plots the spin-up conductance $G_\uparrow$ for model 2a with different asymmetries. It can be seen that the transmission spectra are also composed of the HOMO and LUMO bands for all investigated values of $x$, and the spin-up conductance exhibits several resonant peaks of $G_\uparrow=1$. If the asymmetry factor $x$ becomes larger, the hopping integral along the first helical chain is increased and thus, the bandwidth of the HOMO band broadens. This leads to the shift of the LUMO band toward higher energies, owing to the repulsion effects between the two helical chains. Besides, one notices that the spin-up conductance in the LUMO band can be greater than $e^2/h$ because the system is multi-terminal and its maximum increases with $x$.

Figure~\ref{fig:two}(d) shows the corresponding $\P$ for both the four-terminal molecular device (solid lines) and the eight-terminal one (dotted lines) with various asymmetries. We can see that $\P$ becomes nonzero in the multi-terminal molecular devices, irrespective of the asymmetries we introduced between the two helical chains. In fact, in the multi-terminal set-up, the extra terminals can play similar role as the B\"{u}ttiker's virtual electrode which can cause dephasing. $\P$ is positive (negative) in the LUMO (HOMO) band. Furthermore, the magnitude of $\P$, as well as the energy region of nonzero $\P$, could be significantly increased for both the four-terminal and eight-terminal molecular devices by increasing the asymmetry between the two helical chains. Besides, the spin polarization of the eight-terminal molecular device is larger than that of the four-terminal one [see the solid and dotted lines with identical color of Fig.~\ref{fig:two}(d)], because the spin-filtering effects are enhanced when the system becomes more open.

\begin{figure}[ht]
\includegraphics[width=0.4\textwidth]{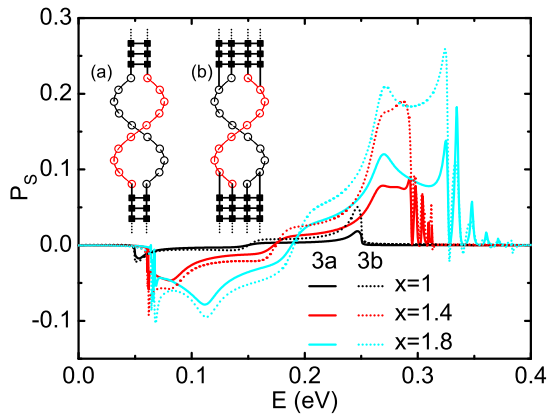}
\caption{\label{fig:three}(Color online) $\P$ for models 3a and 3b with different asymmetries. The inset shows the sketch of the corresponding molecular devices connected to two finite-width electrodes.}
\end{figure}

\section{Two-terminal set-up with finite-width electrodes}

We then investigate the spin transport along two-terminal molecular devices coupled to finite-width electrodes. The insets (a) and (b) of Fig.~\ref{fig:three} sketch the two-leg ladder electrodes and the four-leg ladder ones, respectively, with several sites of the DH molecule connected to each electrode. Here, the electrode Hamiltonian ${\cal H}_{\rm e}^{(D,J)}$ is
\begin{align}
{\cal H}_{\rm e}^{(D,1)} &= \sum_{j=1}^{D}\Big(\sum_{n=-\infty}^{-1}
t_0a_{jn}^\dagger a_{j n+1}^{} \nonumber \\ &+ \!\sum_{n=N+1}^ {+ \infty}\! t_0 a_{jn}^\dagger a_{j n+1}^{}+ {\rm H.c.}\Big) \nonumber \\ &+ \sum_{j=1}^{D-1}\Big(\sum_{n=-\infty}^{0} \lambda_0 a_{jn}^ \dagger a_{j+1 n}^{} \nonumber \\ & + \!\sum_{n=N+1}^{+\infty}\! \lambda_0 a_{jn}^\dagger a_{j+1 n}^{} + {\rm H.c.}\Big)\ , \label{eq8}
\end{align}
with $D=2$ and $4$ for models 3a and 3b, respectively. The interchain hopping integral in the electrode region is considered as $\lambda_0=t_0$.

Although both models 3a and 3b are two-terminal systems, they are obviously more open than models 1a-1d, because more than one site of the molecule is attached to different sites of each electrode in models 3a and 3b. In fact, the connection between the molecule and the electrodes of model 3a (3b) is the same as model 2a (2b), ${\cal H}_{\rm c}^{\rm (3a)}={\cal H}_{\rm c}^{\rm (2a)}$ and ${\cal H}_{\rm c}^{\rm (3b)}={\cal H}_{\rm c}^{\rm (2b)}$. Accordingly, for both models 3a and 3b nonzero $\P$ can be observed in these two systems, as illustrated in the main frame of Fig.~\ref{fig:three}. Similarly, the spin-filtering effects could be enhanced by increasing $x$ for both models. Since the number of sites connected to each electrode in model 3b is twice as much as in model 3a, the spin filter efficiency of the former model is greater than the latter one.

\begin{figure}
\includegraphics[width=0.4\textwidth]{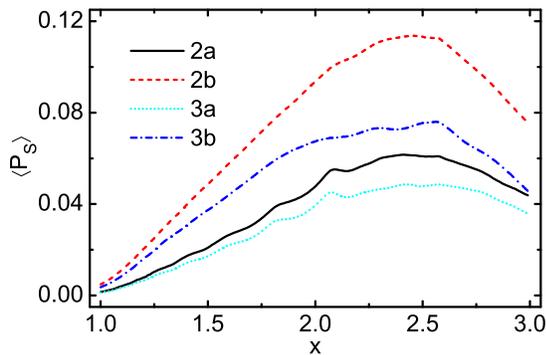}
\caption{\label{fig:four}(Color online) Averaged spin polarization $\langle \P \rangle$ for the molecular devices with various contact configurations, as a function of the asymmetric parameter $x$.}
\end{figure}

In order to further demonstrate the role of the asymmetry on the spin transport along the molecular systems, Fig.~\ref{fig:four} plots the averaged spin polarization $\langle \P \rangle$ versus $x$ with various contact configurations. Here, the averaged spin polarization is defined as
\begin{eqnarray}
\langle \P \rangle =\frac 1 {\Omega}\int_\Omega \P dE,
\end{eqnarray}
with $\Omega$ being the energy range of the LUMO band. It can be seen from Fig.~\ref{fig:four} that the dependence of $\langle \P \rangle $ on $x$ is not monotonic. $\langle \P \rangle $ increases with $x$ in the range $1\leq x \lesssim 2.5$ and is then declined by further increasing $x$. This behavior is related to the intrinsic effects of the DH molecule itself and does not depend upon the particular contact configuration at all. When $x$ is sufficiently large, the hopping integral of the first helical chain is much larger than that of the second one and the electron will be preferentially transmitted along the first helical chain. This is the intermediate status between the spin transport in the DH molecule and in the single-helical molecule. Since an undistorted ssDNA cannot behave as a spin filter, \cite{goehler11,xie,Guo1} the spin polarization will decrease with increasing $x$ in the regime of large $x$.

One may notice that the magnitude of $\langle \P \rangle $ strongly depends on the contact configurations. Indeed, it is determined by both the type of real electrodes and the connection between the molecule and the electrodes. On the one hand, since the number of sites connected to each electrode for models 2b and 3b is twice as large as that for models 2a and 3a, the former two models are more open and exhibit higher $\P$ (see the dashed and dash-dotted lines in Fig.~\ref{fig:four}). On the other hand, model 2a (2b) is switched into model 3a (3b) by coupling two neighboring 1D electrodes with the hopping integral $\lambda_0$ [see Eq.~(\ref{eq8})] and the multi-terminal devices are changed into the two-terminal ones simultaneously. As a result, the openness of the systems is decreased. This can also be understood as follows if we consider models 2a and 3a as an example. In the absence of $\lambda_0$, the two 1D electrodes are separated from each other and the mode number in the electrode region is 2 for model 2a, disregarding the spin degree of freedom. While in the presence of $\lambda_0$, the 1D electrodes are combined together as a whole and the corresponding effective mode number is reduced to 1 when they are coupled to each other extremely tightly with $\lambda_0>2t_0$. In the moderate range of $\lambda_0\in(0,2t_0)$, e.g., $\lambda_0=t_0$ in model 3a, the effective mode number can be 1 or 2 at different energy regions. Similar arguments can be discussed between models 2b and 3b. Accordingly, the spin polarization of model 2a (2b) is higher than model 3a (3b). From the above two points, model 2b possesses the largest $\langle \P \rangle$  (see the dashed line in Fig.~\ref{fig:four}) and model 3a has the smallest $\langle \P \rangle$ (see the dotted line in Fig.~\ref{fig:four}).

\section{Finite-width electrodes with bottleneck}

\begin{figure}[ht]
\includegraphics[width=0.4\textwidth]{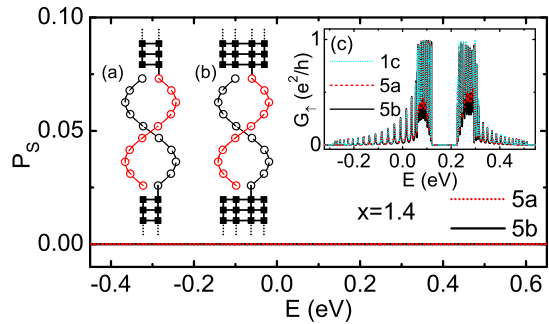}
\caption{\label{fig:five}(Color online) Main frame: $\P$ for models 5a and 5b with $x=1.4$. The insets (a) and (b) display the two-terminal molecular devices with only single site connected to each finite-width electrode. The inset (c) shows $G_\uparrow$  for models 5a and 5b, and also for model 1c as a comparison.}
\end{figure}

Finally, we consider other two-terminal molecular devices with the single site $\{1,1\}$ ($\{2,N\}$) connected to the left (right) finite-width electrode, as illustrated in the insets (a) and (b) of Fig.~\ref{fig:five}. This contact over a single site is a bottleneck in a system of otherwise finite width. These contact configurations are closest to the experiments of the second category discussed in the introduction and the experiment by Xie \textit{et al.}\cite{xie} The electrodes of model 5a (5b) are the same as those of model 3a (3b). Thus, the electrode Hamiltonian ${\cal H}_{\rm e}^{(D,J)}$ is identical to ${\cal H}_{\rm e}^{(D,1)}$ given in \eqref{eq8}. The coupling Hamiltonian $H_{c}$ for models 5a and 5b is
\begin{subequations}
\begin{eqnarray}
{\cal H}_{\rm c}^{\rm (5a)}&=&t_\Left a_{1 0}^\dagger c_{1 1}^{} + t_\Right a_{1N+1}^\dagger c_{2 N}^{} + {\rm H.c.}\ , \label{eq9-1} \\ {\cal H}_{\rm c}^{\rm (5b)}&=&t_\Left a_{2 0}^\dagger c_{1 1}^{} + t_\Right a_{2N+1}^\dagger c_{2 N}^{} + {\rm H.c.}\ . \label{eq9-2}
\end{eqnarray}
\label{eq9}
\end{subequations}

It can be seen from the inset (c) of Fig.~\ref{fig:five} that although the conductance profiles of models 1c, 5a, and 5b are similar, the magnitude of $G_\uparrow$, especially in the energy regions closest to the band gap, is different among the three devices. This is attributed to the quantum interference effects at the interface between the molecule and the electrodes. Since the electrode width is different for models 1c, 5a, and 5b, the injection (ejection) mode is distinct and the conductance will be changed from one  device to another. However, no spin polarization could appear in models 5a and 5b (see the main frame of Fig.~\ref{fig:five}), identical to that observed in model 1c. The physics here is totally different from models 3a and 3b, although the only difference  between models 5a (5b) and 3a (3b) is that in the former case only single site of the molecule is attached to each electrode.

\section{Conclusions}

In summary, we have studied the spin transport along a double-helical molecule by considering various contact configurations in a completely coherent charge transport regime. We find that the conductance and the spin polarization strongly depend upon the contact configurations for the coherent transport case. No spin polarization emerges in the two-terminal set-up when they are coupled to the 1D electrodes or if there is a bottleneck of only one site connecting the molecule to finite-width electrodes. In this case, additional dephasing is necessary to support spin-filtering effects.\cite{Guo1} In contrast, the spin polarization appears in the multi-terminal set-up or the two-terminal one with more than one site of the molecule connected to each finite-width electrode, because in the multi-terminal devices the extra terminals can play similar role as the B\"{u}ttiker's virtual electrode, which can cause dephasing. This effect could be further enhanced by increasing the asymmetry between the two helical chains of the molecule, a result related to that found in Ref.~\onlinecite{Gutierrez13} for a different model, where spin transport through two transport channels on a single helix was studied. The results obtained in the present work are general for any double-helical molecule and reveal that the spin-selective effects could be observed in double-helical molecular devices at low temperature without dephasing by properly tuning the electrode number and the connection between the molecule and the electrodes.

\acknowledgments

This work was supported by the DAAD (54367888), by NBRP of China (2012CB921303), by MINECO (Grants PRI-AIBDE-2011-0.927 and MAT 2010-17180), by NSF-China under Grant No. 11274364, and by PDSF-China under Grant No. 2013M540153. We acknowledge support from the German Excellence Initiative: Cluster of Excellence EXC 1056 ``Center for Advancing Electronics Dresden''~(cfAED).


\begin{thebibliography}{99}

\bibitem{goehler11} B. G\"{o}hler, V. Hamelbeck, T. Z. Markus, M. Kettner, G. F. Hanne, Z. Vager, R. Naaman, and H. Zacharias, Science \textbf{331}, 894 (2011).
\bibitem{xie} Z. Xie, T. Z. Markus, S. R. Cohen, Z. Vager, R. Gutierrez, and R. Naaman, Nano Lett. \textbf{11}, 4652 (2011).
\bibitem{kumar} K. S. Kumar, N. Kantor-Uriel, S. P. Mathew, R. Guliamov, and R. Naaman, Phys. Chem. Chem Phys. \textbf{15}, 18357 (2013).
\bibitem{zach2013} D. Mishra, T. Z. Markus, R. Naaman, M. Kettner, B. Gohler, H. Zacharias, N. Friedman, M. Sheves, C. Fontanesi, Proc. Nat. Acad. Sci. USA \textbf{110}, 14872 (2013).
\bibitem{Guo1} A.-M. Guo and Q.-F. Sun, Phys. Rev. Lett. \textbf{108}, 218102 (2012).
\bibitem{Guo2} A.-M. Guo and Q.-F. Sun, Phys. Rev. B \textbf{86}, 035424 (2012).
\bibitem{Guo3} A.-M. Guo and Q.-F. Sun, Phys. Rev. B \textbf{86}, 115441 (2012).
\bibitem{Gutierrez12} R. Gutierrez, E. D\'{\i}az, R. Naaman, and G. Cuniberti, Phys. Rev. B \textbf{85}, 081404(R) (2012).
\bibitem{Gutierrez13} R. Gutierrez, E. D\'{\i}az, C. Gaul, T. Brumme, F. Dom\'{\i}nguez-Adame, and G. Cuniberti, J. Phys. Chem. C \textbf{117}, 22276 (2013).
\bibitem{yeganeh} S. Yeganeh, M. A. Ratner, E. Medina, and V. Mujica, J. Chem. Phys. \textbf{131}, 014707 (2009).
\bibitem{medina} E. Medina, F. L\'{o}pez, M. A. Ratner, V. Mujica, Europhys. Lett. \textbf{99}, 17006 (2012).

\bibitem{gersten} J. Gersten, K. Kaasbjerg and A. Nitzan, J. Chem. Phys. \textbf{139}, 114111 (2013).
\bibitem{Eremko2013} A. A. Eremko and V. M. Loktev, Phys. Rev. B, \textbf{88}, 165409 (2013).
\bibitem{vager} D. Vager and Z. Vager, Phys. Lett. A \textbf{376}, 1895 (2012).
\bibitem{naaman12} R. Naaman and D. H. Waldeck, J. Phys. Chem. Lett. \textbf{3}, 2178 (2012).
\bibitem{tao1} N. J. Tao, Nature Nanotechnology, \textbf{1}, 173 (2006).
\bibitem{fink}  H. W. Fink and C. Sch\"{o}nenberger, Nature (London) \textbf{398}, 407 (1999).
\bibitem{pablo} P. J. de Pablo, F. Moreno-Herrero, J. Colchero, J. G\'{o}mez Herrero, P. Herrero, A. M. Bar\'{o}, P. Ordej\'{o}n, J. M. Soler, and E. Artacho, Phys. Rev. Lett. \textbf{85}, 4992 (2000).
\bibitem{bezryadin} A. Bezryadin, C. Dekker, and G. Schmid, Appl. Phys. Lett. \textbf{71}, 1273 (1997).
\bibitem{porath} D. Porath, A. Bezryadin, S. de Vries, and C. Dekker, Nature (London) \textbf{403}, 635 (2000).
\bibitem{cui} X. D. Cui, A. Primak, X. Zarate, J. Tomfohr, O. F. Sankey, A. L. Moore, T. A. Moore, D. Gust, G. Harris, and S. M. Lindsay, Science \textbf{294}, 571 (2001).
\bibitem{xu} B. Xu, P. Zhang, X. Li, and N. Tao, Nano Lett. \textbf{4}, 1105 (2004).
\bibitem{cohen} H. Cohen, C. Nogues, R. Naaman, and D. Porath, Proc. Natl. Acad. Sci. U.S.A. \textbf{102}, 11589 (2005).
\bibitem{guo} X. Guo, A. A. Gorodetsky, J. Hone, J. K. Barton, and C. Nuckolls, Nat. Nanotechnol. \textbf{3}, 163 (2008).
\bibitem{grozema} F. C. Grozema, Y. A. Berlin, and L. D. A. Siebbeles, J. Am. Chem. Soc. \textbf{122}, 10903 (2000).
\bibitem{wang} X. F. Wang and T. Chakraborty, Phys. Rev. lett. \textbf{97}, 106602 (2006).


\bibitem{voityuk} A. A. Voityuk, J. Jortner, M. Bixon, and N. R\"{o}sch, J. Chem. Phys. \textbf{114}, 5614 (2001).
\bibitem{senthilkumar} K. Senthilkumar, F. C. Grozema, C. F. Guerra, F. M. Bickelhaupt, F. D. Lewis, Y. A. Berlin, M. A. Ratner, and L. D. A. Siebbeles, J. Am. Chem. Soc. \textbf{127}, 14894 (2005).
\bibitem{hawke} L. G. D. Hawke, G. Kalosakas, and C. Simserides, Eur. Phys. J. E \textbf{32}, 291 (2010).
\bibitem{Ryndyk2009} D. A. Ryndyk, R. Guti\'{e}rrez, B. Song and G. Cuniberti, Springer Series on Chemical Physics, \textbf{93}, 213 (2009).
\bibitem{Lee} D. H. Lee and J. D. Joannopoulos, Phys. Rev. B \textbf{23}, 4997 (1981).
\bibitem{Sun1} Q.-F. Sun and X. C. Xie, Phys. Rev. B \textbf{71}, 155321 (2005).
\bibitem{openness} R. Schuster, E. Buks, M. Heiblum, D. Mahalu, V. Umansky, and H. Shtrikman, Nature (London) \textbf{385}, 417 (1997).
\bibitem{DAmato} J. L. D'Amato and H. M. Pastawski, Phys. Rev. B \textbf{41}, 7411 (1990).


\end{thebibliography}
\end{document}